\documentclass{amsproc}
\usepackage{amscd}
\usepackage{graphics}
\newtheorem{Theorem}{Theorem}[section]
\newtheorem{Proposition}{Proposition}[section]
\newtheorem{Corollary}{Corollary}[section]
\newtheorem{Lemma}{Lemma}[section]

\def\proof{\par{\it Proof}. \ignorespaces}
\def\endproof{{\ \vbox{\hrule\hbox{%
    \vrule height1.3ex\hskip0.8ex\vrule}\hrule }}\par}
\newenvironment{Proof}{\proof}{\endproof}

\theoremstyle{definition}

\newtheorem{Example}[Theorem]{Example}

\theoremstyle{remark}
\newtheorem{Remark}[Theorem]{Remark}

\numberwithin{equation}{section}

\begin{document}

\title{Singular sector of the Burgers-Hopf hierarchy and
deformations of hyperelliptic curves}

\author{Yuji Kodama}
\address{Department of Mathematics, Ohio State University, Columbus,
OH 43210, USA}
\email{kodama@math.ohio-state.edu}

\author{Boris G. Konopelchenko}
\address{Dipartimento di Fisica, Universita di Lecce and Sezione INFN, 73100
Lecce, Italy}
\email{konopel@le.infn.it}

\date{}


\begin{abstract}
We discuss the structure of shock singularities of the Burgers-Hopf
hierarchy. It is shown that the set of singular solutions defines a
stratification of the affine space of the flow parameters in the hierarchy.
The stratification is associated with the Birkhoff decomposition of
the Grassmannian given by the set of linear spaces spanned by the hierarchy.
We then construct integrable hierarchy on each stratum and demonstrate that
it describes a deformation of a hyperelliptic curve parametrizing the stratum.
The hierarchy is called the hidden Burgers-Hopf hierarchy, and we found
the Riemann invarint form and the hodograph solution.
\end{abstract}

\maketitle
\markboth{KODAMA AND KONOPELCHENKO}
   {SINGULAR SECTORS OF THE BURGERS-HOPF HIERARCHY}

\section{Introduction: Singular solutions of the BH hierarchy}
The Burgers-Hopf (BH) hierarchy defined by
\[
\frac{\partial u}{\partial t_n}=c_nu^n\frac{\partial u}{\partial x}, \quad
{\rm with}\quad c_n=(-1)^n\frac{(2n+1)!!}{2^nn!},\quad n=0,1,2,\cdots,
\]
plays an important role in the study of a wide range of phenomena in physics
from hydrodynamics to topological field theory (for examples, see
\cite{whitham:74,witten:90,dubrovin:97}). It is clear that the BH
hierarchy can be obtained by the
dispersionless limit of the KdV hierarchy as well as the dissipationless limit
of the Burgers hierarchy. However the present study does not
require any information of the original hierarchy.

Using the method of
characteristic, the solution of the hierarchy can be expressed by the
hodograph form with $t_0\equiv x$,
\begin{equation}
\label{hodograph}
\Omega(u,x,t_1,t_2,\cdots):=x+\sum_{j=1}^{\infty}c_ju^jt_j-f(u)=0,
\end{equation}
where $f(u)$ may be determined by the initial data, e.g.
$u(x,0,0,\cdots)=f^{-1}(x)$ for a monotone data. Then the BH hierarchy gives
a deformation of the data $x=f(u)$ in each $t_j$-direction, and this defines
an infinite dimensional surface given by (\ref{hodograph}).

Here we are interested in the class of solutions of the BH hierarchy
with singularity of shock wave type.
The singularity of this type
corresponds to a point on the surface where the derivative $\partial
u/\partial x$ blows up at this point, that is, the equation
(\ref{hodograph})
is not invertible for $u$. This condition is equivalent to
$\Omega':=\partial\Omega/\partial u=0$ at the point, which implies that the
solution $u$ can be expressed as a function of $(t_1,t_2,\cdots)$.
Thus the conditions $\Omega=0$ and $\Omega'=0$ define a
singular sector of codimension
$\ge 1$ in ${\mathbb C}^{\infty}$. The affine space ${\mathbb C}^{\infty}$
is then stratified with those singular sectors
as follows: Let us first defined the solution sets,
\[
U_n:=\left\{~u~\Big|~\Omega^{(j)}=0~ (0\le j\le n),~\Omega^{(n+1)}\ne
0~\right\},
\]
where $\Omega^{(i)}=\partial^i\Omega/\partial u^i$. Then
the singular sector of codimension $m$ can be defined as
\begin{equation}
\label{strataZ}
Z_m:=\left\{~(t_0,~t_1,~t_2,~\cdots~)\in
{\mathbb C}^{\infty}~\Big|~u\in U_j ~{\rm such~that}~ m=\sum_{j=1}^{\infty}
j|U_j|.~\right\}.
\end{equation}
where $|U_j|$ is the number of solutions in $U_j$.
For examples, $Z_0$ consists of the points where $\Omega(u)=0$ has only simple
roots, and the points in $Z_1$ correspond to the case where $\Omega(u)=0$ and
$\Omega'(u)=0$ has just only one common root and no roots for
$\Omega^{(j)}(u)=0,~j\ge 2$.
The closure of the set of those singular sectors then forms a 
decomposition of the affine
space
${\mathbb C}^{\infty}$, i.e.
\[
{\mathbb C}^{\infty}=\overline{\bigcup_{m\ge 0}Z_m}, \quad {\rm 
disjoint~union}.
\]
In order to show some explicit structure of the set $Z_m$, let us consider
the case where $f(u)$ is given by a $(N+1)$th degree of polynomial in $u$.
Then on  a $(N+1)$ dimensional subspace of ${\mathbb C}^{\infty}$ with
$t_n=0$ for $n\ge N+1$, the equation (\ref{hodograph}) can be written by
\begin{equation}
\label{An}
\Omega_{N+1}(u,x,t_1,\cdots,t_N)=x+\sum_{j=1}^{N}c_ju^jt_j=u^{N+1}+\sum_{j=0
}^{N}b_ju^j.
\end{equation}
Since all the coefficients $b_j$'s on the right hand side can be absorbed by
shifting the times, $(x-b_0,t_1-(b_1/c_1),t_2-(b_2/c_2),\cdots)$, we just
set
all $b_j=0$. Then we consider the singular sectors, denoted as $Z_m^{N+1}$,
on the subspace ${\mathbb C}^{N+1}$. The equation (\ref{An}) is well known as a
universal unfolding of a singularity of $A_N$-type \cite{arnold:96}.
The singularity of this type includes cusp, swallow tail and butterfly
for $N=2,3$ and $4$. As an example we just consider
the case with $N=2$.
\begin{Example}
The hodograph solution (\ref{An}) with $N=2$ is given by
\[
\Omega_3(u,x,t_1,t_2)=x-\frac{3}{2}ut_1+\frac{15}{8}u^2t_2-u^3=0.
\]
Then the singular sector $Z^3_1$ is given by
\[Z^3_1=Z^{3+}_1\cup Z^{3-}_1, \quad {\rm with} \quad
Z^{3\pm}_1=\left\{~(x,t_1,t_2)~\Big|~\Omega_3(u^{\pm},x,t_1,t_2)=0,~
u^+\ne u^-~\right\},
\]
where $u^{\pm}$ are the roots of $\Omega_3'=0$, i.e.
\[
u^{\pm}=\frac{1}{8}\left(5t_2\pm \sqrt{25t^2_2-32t_1}\right).
\]
The sets $Z^{3\pm}_1$ obviously intersects on the double points of $u^+=u^-$,
and forms a cusp singularity.
The intersection provides the higher singular sector $Z^3_2$ which is given by
a twisted cubic curve,
\[
Z^3_2=\left\{~(x,t_1,t_2)~\Big|~x=u^3,~t_1=2u^2,~t_2=(8/5)u~\right\}.
\]
The real sections of those sectors will be an interesting object to study.
\end{Example}
\begin{Remark}
A stratification of certain (compact) manifold is an interesting problem of
algebraic geometry, and has several intimate connections with the study of
integrable systems. For examples, the isospectral manifolds of the periodic
Toda lattices are given by Jacobian varieties which have the stratification
based on the theta divisor \cite{adler:91,adler:93,kodama:02}.
Also in the case of finite nonperiodic Toda lattices, the isospectral manifolds
are compactified in the flag manifolds, and are decomposed into
  the Bruhat cells which correspond to the sets of singular solutions
\cite{flaschka:91,casian:02}.
(See also \cite{vanhaecke:01} for several examples of the stratifications
of Jacobian varieties related to finite dimensional hamiltonian systems.)
The present study will provide an example of the stratification
related to integrable systems of hydrodynamic type.
\end{Remark}

In this paper, we construct an integrable hierarchy defined on each singular
sector $Z_m$. Those hierarchies can be also obtained by the dispersionless
limits
of the hidden hierarchies of the KdV equation considered in \cite{manas:97}.
We call those the hidden BH hierarchies, and we construct those as a
regularization of the
shock singularity of the BH hierarchy. Each hierarchy on $Z_m$ is called the
${\rm BH}_m$ hierarchy. In Section 2, we discuss the Grassmannian structure
of the singular sectors and show the connection to hyperelliptic curves.
Then in Section 3, we define the ${\rm BH}_m$ hierarchy on
$Z_m$ as a deformation of a hyperelliptic curve of genus $m$. We also
show that the hierarchy can be put into the Riemann invariant form, and then
construct the
hodograph solution. The Riemann invariant form has been also obtained
in a different approach in \cite{ferapontov:91}. In Section 4, we give a
brief discussion on several
extensions of the present construction of the hidden dispersionless
hierarchies. We also mention some related topics such as a topological field
theory \cite{aoyama:96} and a possible
stratification of the Frobenius manifold as obtained by those hidden
hierarchies \cite{kodama:98}.

\section{Grassmannian structure of the BH hierarchy}
In this section we discuss the Grassmannian structure of the solutions
of the BH hierarchy.
Let us first recall (see for example \cite{kodama:89}) that the BH hierarchy
can be formulated as
\begin{equation}
\label{dkdv}
\frac{\partial p}{\partial t_j}=\frac{\partial}{\partial x}Q_j,
\quad {\rm with}\quad Q_j=(\lambda^{2j+1})_{+p},~ j=0,1,2,\cdots,
\end{equation}
where $(\lambda^n)_{+p}$ indicates the polynomial part of $\lambda^n$ in $p$
having the algebraic relation,
\begin{equation}
\label{genus0}
p^2=\lambda^2+u.
\end{equation}
The hierarchy (\ref{dkdv}) can be considered as a deformation of the curve
(\ref{genus0}) which corresponds to the Riemann surface of genus 0 (sphere)
with the compactification. Later we will use $k=\lambda^2$, and then we
assign the degrees as
\[
{\rm Deg}(k)=2, \quad {\rm Deg}(p)=1.
\]
One introduces a function $S(x,t_1,t_2,\cdots)$ which plays an important role
in the theory of dispersionless hierarchy. The $S$ function is defined by
rewriting (\ref{dkdv}) in the form,
\begin{equation}
\label{sflow}
\frac{\partial S}{\partial t_j}=Q_j, \quad{\rm for} \quad j=0,1,2,\cdots.
\end{equation}
Since $Q_j=(\lambda^{2j+1})_{+p}$ is a polynomial of degree $2j+1$ in $p$
and $p^2=k+u,~k=\lambda^2$, the $S$ function can be expressed by a Laurent
series in ${\mathbb C}[[k,k^{-1}]]\cdot\sqrt{k}$,
\[
S(t_0,t_1,t_2,\cdots)=\sum_{i=0}^{\infty}k^{i+\frac{1}{2}}t_i+
\sum_{j=0}^{\infty}
\frac{1}{k^{j+\frac{1}{2}}}F_j(t_0,t_1,t_2,\cdots).
\]

To characterize the solution space of the BH hierarchy, we now
define a linear space spanned by the flows (\ref{sflow}) in a manner similar
to the case of the KP hierarchy \cite{adler:94},
\[\begin{array}{lll}
W_0&:=& {\rm Span}_{\mathbb C}\{~S_{t_0},~S_{t_1},~S_{t_2},~\cdots~\}\\
  &=&{\rm Span}_{\mathbb C}\{~Q_0,~Q_1,~Q_2,~\cdots~\}
  \end{array}
  \]
Since $Q_i$ is a polynomial in $p$ and $p^2=k+u$,
one can also consider $Q_i$ be an element in ${\mathbb C}[k]\cdot p$.
This observation will be crucial for defining an integrable hierarchy on
the singular sector. Thus we have
\[
W_0={\rm Span}_{\mathbb C}\{~p,~kp,~k^2p,~\cdots~\}\cong {\mathbb C}[k]\cdot p.
\]
This implies the inclusion,
\[
kW_0\subset W_0,
\]
which is the condition for the KdV reduction in terms of the KP hierarchy
\cite{pressley:86}.

One should however note that the linear space $W_0$ cannot be defined on the
singular sectors where the flows (\ref{dkdv}) blow up and the commutativity
among the $S$ flows is not defined. Recall that the conditions on the
hodograph solution $\Omega=0$ provide a constraint on the coordinates. For
example,
the first condition  $\Omega'=0$ gives
\[
c_1t_1+\sum_{j=2}^{\infty}jc_ju^{j-1}t_j=f'(u)
\]
which implies that $u$ can be determined as a function of $t_j$ for $j\ge 1$.
Then with $\Omega=0$, the coordinate $x$ can be considered as a function of
the other times. Hence one can expect to have a map $\psi^m : {\mathbb
C}^{\infty}\longrightarrow  Z_{m}$ so that
\[
  \psi^m(t_m,t_{m+1},t_{m+2},\cdots)
=(a_0,a_1,\cdots,a_{m-1},t_m,t_{m+1},\cdots),
  \]
where $a_i,~0\le i\le m-1$ are some functions of $t_j,~ j\ge m$.
The map can be defined at least locally on the sector
$Z_m$, and it defines a patch on $Z_m$. The $S$-function then takes the form,
\begin{equation}
\label{sm}
S^m(t_m,t_{m+1},\cdots)=k^{m+\frac{1}{2}}\sum_{i=0}^{\infty}k^{i}t_{m+i}
+ \sum_{i=0}^{m-1}a_i(t_m,t_{m+1},\cdots)k^{i+\frac{1}{2}}
+O(k^{-{\frac{1}{2}}}).
\end{equation}
  The linear space spanned by the flows $\partial S^m/\partial t_j~j\ge m$
under this restriction on the times defines
  \[
W_m={\rm Span}_{\mathbb
C}\{~S^m_{t_m},~S^{m}_{t_{m+1}},~S^m_{t_{m+2}}~\cdots~ \}\cong k^mW_0.
\]
Then introducing new variable $p_m:=S^m_{t_m}$, we see $W_m$
as the polynomial ring,
\begin{equation}
\label{mring}
W_m\cong {\mathbb C}[k]\cdot p_m, \quad {\rm as~a~vector~space}.
\end{equation}
The Grassmannian $Gr$ is a set of all linear spaces $W_m,~ m\ge 0$, and it has
the Birkhoff decomposition \cite{pressley:86},
\[
Gr=\bigcup_{m\ge 0}\Sigma_m,
\]
where the Birkhoff stratum $\Sigma_m$ is the set of all linear spaces of $W_m$.
The structure of the Grassmannian is then the same as the KdV hierarchy by
making the identification with $k=\lambda^2$,
\[
W_m\cong \left\{\lambda^{-m},\lambda^{-m+2},\cdots,\lambda^{m-2},\lambda^{m},
\lambda^{m+1},\cdots\right\},
\]
where $\lambda^2 W_m\subset W_m$ (i.e. $kW_m\subset W_m$) \cite{adler:94}.
The BH hierarchy is thus defined on the
principal stratum $\Sigma_0$.

Since the degrees of $k$ and $p_m$ are assigned as
\[
{\rm Deg}(k)=2,\quad {\rm Deg}(p_m)=2m+1,
\]
one can consider a plane curve ${\mathcal C}_m$ defined by
\begin{equation}
\label{curve}
{\mathcal C}_m:=\left\{~(k,p_m)\in {\mathbb
C}^2~\Big|~p^2_m=k^{2m+1}+\sum_{i=0}^{2m}u_ik^{2m-i}~\right\},
\end{equation}
which defines a hyperelliptic curve of genus $m$ and is a smooth affine
variety for the generic values of $u_i$'s, the deformation
parameters.
Then the ring $W_m={\mathbb C}[k]\cdot p_m$ can be considered as a part of
the quotient ring ${\mathbb C}[k,p]/{\mathcal C}_m$ with the curve
${\mathcal C}_m$. One should also note that there is a natural reduction
of the curve,
\[
{\mathcal C}_m \longrightarrow {\mathcal C}_{m-1} \quad
(p_m =kp_{m-1}),
\]
which corresponds to the conditions on the parameters, $u_{2m}=u_{2m-1}=0$.
This can be considered as a dispersionless analog of the regularization of
the singularity
in the KP hierarchy by the B\"acklund transformation \cite{adler:94}.
Thus the stratum $\Sigma_m$ is parametrized by the curve ${\mathcal C}_m$,
and the boundary of the stratum corresponds to a singular curve of degenerate
genus $m$. The higher order stratum is obtained by the desingularization of
the curve by increasing the genus.
The curve ${\mathcal C}_m$
can be obtained by the dispersionless limit of the hidden KdV hierarchy
considered in \cite{manas:97} and also by a quasi-classical $\bar\partial$
dressing method introduced in \cite{konopelchenko:01}.
In the next section we will consider
a deformation of the curve by defining a system of equations for the
parameters $u_i$ in  ${\mathcal C}_m$ where the deformation is
parametrized by the times $t_j$ for $j\ge m$.

\section{The hidden BH hierarchies}
We now construct an integrable deformation of the curve ${\mathcal C}_m$
in (\ref{curve}), which is
defined on the singular sector $Z_m$. First recall the quotient ring of
${\mathbb C}[k,p_m]$ over an ideal ${\mathcal C}_m$ which has a split,
\[
\displaystyle{{\mathcal R_m}:=\frac{{\mathbb C}[k,p_m]}{{\mathcal
C_m}}={\mathbb C}[k]\oplus
{\mathbb C}[k]\cdot p_m.}
\]
Then a deformation of the curve ${\mathcal C}_m$ with an infinite number of
  deformation parameters is defined in the same form as (\ref{sflow}),
\begin{equation}
\label{smflow}
\frac{\partial S^m}{\partial t_j}=Q^m_j, \quad Q_j^m\in {\mathbb C}[k]\cdot
p_m,
\quad j=0,1,2,\cdots,
\end{equation}
where $S^m$ is given by (\ref{sm}), and ${\rm Deg}(Q_j^m)=2(m+j)+1$ with
$Q_0^m=p_m$. With this definition, the functions $a_i$'s in $S^m$ can be
explicitely written in terms of $u_i$'s in the curve ${\mathcal C}_m$. Here
we have relabeled the times as $t_{m+i}\to t_i$, but we do not think it will
cause any confusion.
As we will show below, all the flows in (\ref{smflow}) are compatible
for appropriate form of $Q_j^m$, that is, $\partial Q_i^m/\partial
t_j=\partial Q_j^m/\partial t_i$, and in particular we have for the case
$i=0$,
\begin{equation}
\label{pmflow}
\frac{\partial p_m}{\partial t_j}=\frac{\partial}{\partial x}Q_j^m,
\end{equation}
where we denote $x=t_0$. This provides the system of equations of
hydrodynamic type for $u_i$. Also notice that the flows (\ref{pmflow}) are
compatible with
the automorphism of the curve, $p_m\to -p_m$.

Now we show the compatibility of the flows in (\ref{smflow}).
We first note
\begin{Lemma}
The $Q_i^m$'s in (\ref{smflow}) are given by
\[
Q_i^m=\left(\frac{k^{m+i+\frac{1}{2}}}{p_m}\right)_{+}p_m,\quad {\rm
for}\quad i=0,1,2,\cdots,
\]
where $(\cdot)_+$ denotes the projection onto ${\mathbb C}[k]$.
\end{Lemma}
\begin{Proof}
Since $\partial p_m/\partial t_i\sim O(k^{m-\frac{1}{2}})$, one can set
\[
Q_i^m \sim k^{m+i+\frac{1}{2}} +O(k^{m-\frac{1}{2}}).
\]
This implies
\[
\frac{Q_i^m}{p_m}\sim \frac{k^{m+i+\frac{1}{2}}}{p_m}+O(k^{-1}).
\]
The statement of the Lemma then follows from $Q_i^m/p_m\in {\mathbb C}[k]$.
\end{Proof}
This formula is a non-zero genus extension of the case with $m=0$. In fact
we have
\begin{Corollary}
\[
Q_i =(k^{i+\frac{1}{2}})_{+p}=\left(\frac{k^{i+\frac{1}{2}}}{p}\right)_{+}p,
\quad {\rm with}\quad
p^2=k+u,
\]
where $Q_i$ is defined in (\ref{dkdv}) with $k=\lambda^2$.
\end{Corollary}
\begin{Proof}
We first note
\[
\left(\frac{k^{i+\frac{1}{2}}}{p}\right)_+p=k^{i+\frac{1}{2}}-
\left(\frac{k^{i+\frac{1}{2}}}{p}\right)_-p,
\]
where $(.)_-$ is the non-polynomial part in $k$. Since ${\rm Deg}(k)=2$ and
${\rm Deg}(p)=1$, we have
\[
\left(\frac{k^{i+\frac{1}{2}}}{p}\right)_-p=\left(\left(\frac{k^{i+\frac{1}{
2}}}{p}\right)_-p
\right)_{-p}=\left(k^{i+\frac{1}{2}}\right)_{-p}.
\]
This implies the result.
\end{Proof}
In order to show the compatibility of the flows (\ref{smflow}), we note
\begin{Lemma}
The flows (\ref{smflow}) can be put into the Lax form,
\begin{equation}
\label{kflow}
\frac{\partial k}{\partial t_i}=\{Q_i^m,k\}:=\frac{\partial Q_i^m}{\partial
p_m}
\frac{\partial k}{\partial x}-\frac{\partial Q_i^m}{\partial x}
\frac{\partial k}{\partial p_m},
\end{equation}
where $p_m$ is then considered to be a constant parameter.
\end{Lemma}
\begin{Proof}
The flows $\partial p_m/\partial t_i=\partial Q_i^m/\partial x$ in
(\ref{smflow})
with $x=t_0$ can be written in terms of the differential three forms,
\[
dp\wedge dk\wedge dx=dQ_i^m\wedge dt_i\wedge dk.
\]
Then assuming $k=k(p,x,t)$, we obtain the result.
\end{Proof}
It is standard that the following Proposition proves the compatibility of
the flows, that is, $\partial^2 k/\partial t_i\partial t_j=
\partial^2 k/\partial t_j\partial t_i$:
\begin{Proposition}
\begin{equation}
\label{compatibility}
\frac{\partial Q_i^m}{\partial t_j}-\frac{\partial Q_j^m}{\partial t_i}
+\{Q_i^m,Q_j^m\}=0.
\end{equation}
\end{Proposition}
\begin{Proof}
Note that writing $Q_i^m=\beta^+_i p_m$ with $\beta^+_i\in{\mathbb C}[k]$,
the equation (\ref{compatibility})
can be
expressed by the following equation for $\beta^+_i$'s,
\[
\frac{\partial \beta^+_i}{\partial t_j}-\frac{\partial \beta^+_j}{\partial
t_i}+
\{\{\beta^+_i,\beta^+_j\}\}=0,
\]
where
\[
\{\{f,g\}\}:=f\frac{\partial g}{\partial x}-g\frac{\partial f}{\partial x}
+p_m\{f,g\}.
\]
This equation can be proven as follows: First note
\[
\frac{\partial\beta_i}{\partial t_j}=\{\{\beta^+_j,\beta_i\}\}, \quad {\rm
for}\quad
\beta_i=\frac{k^{m+i+\frac{1}{2}}}{p_m}, \quad (\beta_j^+=(\beta_j)_+).
\]
With the decomposition $\beta_i=\beta^+_i+\beta^-_i$, we have
\[
\frac{\partial\beta_i^+}{\partial t_j}+\frac{\partial\beta_i^-}{\partial
t_j}=\{\{\beta^+_j,\beta_i^+\}\}+\{\{\beta^+_j,\beta_i^-\}\}.
\]
We also have
\[
\frac{\partial\beta_j^+}{\partial t_i}+ \frac{\partial\beta_j^-}{\partial
t_i}=-\{\{\beta^-_i,\beta_j^+\}\}-\{\{\beta^-_i,\beta^-_j\}\},
\]
where we have used $\{\{\beta_i,\beta_j\}\}=0$. Combining those two
equations and projecting on ${\mathbb C}[k]$, we obtain the compatibility
equation for
$\beta^+_i$ as well as $\beta^-_i$. This completes the proof.
\end{Proof}
The simplest example of the ${\rm BH}_m$ equations in terms of $u_i$'s is
given by the $t_1$-flow of the ${\rm BH}_1$ equation (see Remark \ref{general}
below for the general form),
\[
\displaystyle{\frac{\partial}{\partial t_1}
\left(\begin{matrix}
u_0 \\ u_1 \\ u_2 \end{matrix}\right)=
\left(\begin{matrix}
-\frac{3}{2}u_0 & 1 & 0 \\
-u_1   &  -\frac{1}{2} u_0 & 1 \\
-u_2   &  0       & -\frac{1}{2}u_0
\end{matrix}\right) \frac{\partial}{\partial x}\left(\begin{matrix}
u_0 \\ u_1 \\ u_2 \end{matrix}\right).}
\]
As a consequence of the Lax form (\ref{kflow}), we have
\begin{Corollary}
The ${\rm BH}_m$ hierarchy (\ref{kflow}) can be put into the Riemann invariant
form,
\begin{equation}
\label{riemann}
\frac{\partial \kappa_i}{\partial t_j}=\phi^j_i\frac{\partial
\kappa_i}{\partial x}, \quad {\rm for}\quad i=0,1,2,\cdots,2m,
\end{equation}
where the Riemann invariants $\kappa_i$'s are the roots of the polynomial
associated with the curve ${\mathcal C}_m$, and $\phi^j_i$ is given by the
$(\partial Q_j^m/\partial p_m)(k)$ at $k=\kappa_i$, that is,
\[
p^2_m=\prod_{i=0}^{2m}(k-\kappa_i), \quad
{\rm and}\quad
\phi^j_i=\phi^j(\kappa_i)=\left(\frac{k^{j+m+\frac{1}{2}}}{p}\right)_+\Big|_
{k=\kappa_i}.
\]
\end{Corollary}
\begin{Proof}
First note that from the curve $p_m^2=F(k)$ we have
\[
2p_m=\frac{\partial F}{\partial k}\cdot\frac{\partial k}{\partial p_m},
\]
  from which ${\partial k}/{\partial p_m}=0$ implies $p_m=0$. Then the
  evaluation of (\ref{kflow}) at the roots of $p_m^2=0$ completes the proof.
\end{Proof}
For example, the Riemann invariant form for the first flow of the ${\rm
BH}_m$ equation
is given by
\[\displaystyle{\frac{\partial \kappa_i}{\partial t_1}=
\left(\kappa_i-\frac{1}{2}u_0 \right)\frac{\partial \kappa_i}{\partial x},
\quad
i=0,1,\cdots,2m,}
\]
where $ u_0=-\sum_{i=0}^{2m}\kappa_i$.
The Riemann invariant form has been obtained in \cite{ferapontov:91}.
We here note that the Riemann invariant form has an obvious reduction with
$\kappa_j=0$ for some $j$'s. This corresponds to the reduction of the
genus as mentioned in the end of the previous section. There is also
another reduction with all $\kappa_0=\cdots=\kappa_{2m}$ which is associated to
a degenerate case of ${\mathcal C}_m$ equivalent to $p_m^2=k^{2m+1}$ of
genus $0$.

With the form (\ref{riemann}), we can construct the hodograph solution of
the hierarchy as shown in \cite{kodama:89}.
We first remark that the functions $\phi^j(k)$ in (\ref{riemann}) form a finite
ring of dimension $2m+1$,
\begin{equation}
\label{F}
{\mathcal F}_m:=\frac{{\mathbb C}[k]}{p_m^2}=
{\rm Span}_{\mathbb C}\left\{\phi^0,\phi^1,\cdots,\phi^{2m}\right\}.
\end{equation}
Note here that $\phi^j(k)$ is a monic polynomial of ${\rm Deg}(\phi^j)=2j$ with
${\rm Deg}(k)=2$ and $\phi^0=1$. Then one can write the higher $\phi^{2m+j}$
for $j\ge 1$ in
the form,
\[
\phi^{2m+j}=\sum_{i=0}^{2m}\mu_i^j(\kappa)\phi^i \quad {\rm mod}~ p_m^2,
\]
where $\mu_i^j(\kappa)$ are functions of the roots $\kappa_i$ of $p_m^2=0$.
We also note that the $\mu_i^j(\kappa)$'s are functions of the symmetric
polynomials $\nu_i$ of the roots $\kappa_j$,
which are given by $\nu_i=(-1)^iu_{i-1},~i=1,\cdots,2m+1$ with
\[
\nu_1=\sum_{j=0}^{2m}\kappa_j,\quad \nu_2=\sum_{i<j}\kappa_i\kappa_j,
\quad \nu_3=\sum_{i<j<l}\kappa_i\kappa_j\kappa_l\quad \cdots\quad
\nu_{2m+1}=\prod_{i=0}^{2m}\kappa_i.
\]
For example, we have for $j=1$
\[
\mu^1_{2m}=-\frac{3}{2}u_0,\quad \mu^1_{2m-1}=-\frac{3}{2}u_1-\frac{3}{8}u_0
^2,\quad
\mu^1_{2m-2}=-\frac{3}{2}u_2-\frac{3}{4}u_0u_1+\frac{1}{16}u_0^3,~\cdots.
\]
Then we have
\[
\frac{\partial \kappa_l}{\partial t_{2m+j}}=\sum_{i=0}^{2m}
\mu_i^j(\kappa)\frac{\partial \kappa_l}{\partial t_i}, \quad l=0,1,\cdots, 2m,
\]
which implies that all the roots $\kappa_l$ are constants along the
characteristic curve,
\[
\frac{dt_0}{\mu_0^j(\kappa)}=\cdots=\frac{dt_{2m}}{\mu_{2m}^j(\kappa)}=\frac
{dt_{2m+j}}{-1},
\quad {\rm for~all}~~ j\ge 1.
\]
The integration of this equation gives
\begin{equation}
\label{string}
t^0_i=t_i+\sum_{j=1}^{\infty}\mu_i^j(\kappa)t_{2m+j}, \quad i=0,1,\cdots, 2m,
\end{equation}
where $t^0_i$ is the initial position of the characteristic at
$t_{2m+j}=0,~j\ge 1.$  As a simplest example with $m=1$, we have an explicit
solution from (\ref{string}) by setting $t_{2m+j}=(3/2)\delta_{j1}$ and
$t^0_i=0$,
\[
u_0=t_2,\quad u_1=t_1-\frac{1}{4}t_2^2,\quad
u_2=x-\frac{1}{2}t_1t_2+\frac{1}{6}t_2^3.
\]
Also taking the sum $\sum_{i=0}^{2m}\phi^it_i^0$ of
(\ref{string}), we obtain
\begin{Theorem}
The hodograph solution of the ${\rm BH}_m$ hierarchy is given by
\begin{equation}
\label{mhodograph}
x+\sum_{i=0}^{\infty}\phi^i(\kappa_j)t_i=0,\quad j=0,1,\cdots,2m,
\end{equation}
where the times $(x,t_1,t_2,\cdots)$ can be shifted arbitrary.
\end{Theorem}
This formula will be useful to study the singular structure of the ${\rm
BH}_m$ hierarchy as in the case of the BH hierarchy.
We will report a detailed analysis elsewhere.

\begin{Remark}
\label{general}
The general form of the ${\rm BH}_m$ hierarchy can be written in the
following form with the polynomial $\phi^i(k)$ for $k$ being replaced by a
matrix $K$,
\[
\frac{\partial U}{\partial t_i}=\phi^i(K)\frac{\partial U}{\partial x},
\quad {\rm with}\quad U=(u_0,u_1,\cdots,u_{2m})^T,
\]
where $K$ is the (companion) matrix given by
\[
K=\left(
\begin{matrix}
-u_0 & 1  & \cdots & \cdots & 0\\
-u_1 & 0  & \cdots & \cdots & 0\\
\vdots& \vdots& \ddots &\ddots & \vdots\\
-u_{2m-1}&0 &\cdots &\cdots & 1\\
-u_{2m}& 0 & \cdots & \cdots& 0
\end{matrix}
\right).
\]
For example, $\phi^1(K)=K-(1/2)u_0I$ with $I=(2m+1)\times(2m+1)$ 
identity matrix.
\end{Remark}

\section{Further extensions and discussion}
The present construction of the hidden hierarchies can be extended to the case
with the hyperelliptic curve associated with an {\it even} degree polynomial,
\[
\tilde{\mathcal C}_m:=\left\{~(k,p)\in {\mathbb
C}^2~\Big|~p^2=k^{2m+2}+\sum_{i=1}^{2m+2}
v_ik^{2m+2-i}~\right\}.
\]
Here the degree of $k$ is assigned as ${\rm Deg}(k)=1$ so that ${\rm
Deg}(p)=m+1$.
The hierarchy of deformations of the curve $\tilde{\mathcal C}_m$ can be also
obtained as the dispersionless limit of the integrable hierarchy associated
with
the energy-dependent Schr\"odinger potentials considered in \cite{martinez:80},
which coincides with the hidden hierarchy for the Jaulent-Miodek hierarchy
\cite{jaulent:76} (see also \cite{konopelchenko:99}).
  We here call the integrable hierarchy associated with the curve $\tilde
{\mathcal C}_m$ the ${\rm dJM}_m$ hierarchy.
Most of the results obtained in Sections 2 and 3 remains the same for this
case. In particular, the singular sectors $Z_m$ for the ${\rm dJM}_0$
hierarchy
are given by the higher order intersections of two curves given by the
hodograph
solution for $v_1$ and $v_2$, which are defined as follows: Let the curve
$\tilde{\mathcal C}_0$ be given by
\[
p^2=k^2+v_1k+v_2=(k-\kappa_1)(k-\kappa_2).
\]
The hodograph solution has the same form as (\ref{hodograph}),
\[
\Omega^j:=x+\sum_{i=0}^{\infty}\phi^i(\kappa_j)t_i=0, \quad j=1,2.
\]
Then the singular sectors are defined by
\[ {Z}_m =\left\{ (t_0,t_1,t_2,\cdots)\in {\mathbb C}^{\infty}~\Big|
~(\kappa_1,\kappa_2)\in U_{j},~~ m=\sum_{j=1}^{\infty}j|U_j|~\right\},
\]
where $U_j$ are defined by
\[
U_j=\left\{ ~(\kappa_1, \kappa_2)~ \Big|~ \Omega^1=\Omega^2=0,
~\tilde \Omega_{(i)}=0~(1\le i\le j),~\tilde\Omega_{(j+1)}\ne 0~\right\}.
\]
Here the function $\tilde\Omega_{(i)}(\kappa_1,\kappa_2)$ may be given by
$\tilde \Omega_{(i)}=D|_{\omega=\Omega^1_2/\Omega^1_1}$ with,
\[
D:=\left|\begin{matrix}
(\Omega^1_1)^i &(\Omega^2_1)^i\\
\Omega^1_{(i)} &\Omega^2_{(i)}
\end{matrix}\right|, \quad \Omega^j_{(i)}=\left(\frac{\partial }{\partial
\kappa_2}
-{\omega}(\kappa_1,\kappa_2)\frac{\partial}{\partial\kappa_1}\right)^i\Omega^j,
\]
where $\Omega^i_j=\partial \Omega^i/\partial \kappa_j$.
Note that $\tilde \Omega_{(1)}$ is just the Jacobian determinant. The singular
structure of the dispersionless JM hiearchy is then similar to the case of
the BH hierarchy. We will report a detailed discussion on the stratification
determined by the ${\rm dJM}_m$ hierarchies elsewhere.

One should also note that the integrable deformation of the algebraic
curve defined by
either ${\rm BH}_m$ or ${\rm dJM}_m$ hierarchy is rather a general property
of the
dispersionless integrable hierarchy. For example, the dispersionless limit
of the first hidden KP hierarchy discussed in \cite{konopelchenko:00} leads to
the standard form of the elliptic curve,
\[
{\mathcal C}_{KP}~:~ p^2+v_0pq+v_1p=q^3+u_0q^2+u_1q+u_2,
\]
and the corresponding deformation has the form,
\[\left\{\begin{array}{lll}
\displaystyle{\frac{\partial p}{\partial
t_n}}&=&\displaystyle{\frac{\partial}{\partial y}Q_n,}\\
\displaystyle{\frac{\partial q}{\partial
t_n}}&=&\displaystyle{\frac{\partial}{\partial x}Q_n,}
\end{array}
\right.\quad {\rm with}\quad \displaystyle{
Q_n\in\frac{{\mathbb C}[p,q]}{{\mathcal C}_{KP}}, \quad n=0,1,2,\cdots,}
\]
where ${\rm Deg}(Q_n)=n+2$ with ${\rm Deg}(p)=3,~{\rm Deg}(q)=2$ and
$t_0=x,~t_1=y$.
This contains the ${\rm BH}_1$ hierarchy as the reduction with $q=k=$constant
and $v_0=v_1=0$, and the hidden dispersioless Boussinesq hierarchy with the
reduction
$p=k=$constant and $u_0=0$. The singular structures determined by the hidden
dispersionless KP hierarchy
and its reductions (such as the Gel'fand-Dikii type) will be discussed
elsewhere. We also remark that the integrable deformation of hyperelliptic
curve discussed in the present paper is different from that given by the
Whitham
hierarchy (see for example \cite{flaschka:80}). The Whitham hierarchy describes
a slow modulation over a quasi-periodic solution of the original dispersive
equation, and is obtained as a dispersive regularization of the dispersionless
hierarchy. However our deformation describes each stratum (singular sector)
of the Birkhoff decomposition, which is parametrized by the hyperelliptic
curve, and it has no direct connection with the original dispersive
equation.

As a final remark, we would like to point out that the ${\rm BH}_m$
hierarchy may not
have a single free energy (or prepotential) which plays a central role
of a topological field theory. In the dispersionless KP hierarchy,
the free energy can be obtained by integrating twice the function
$G_{ij}$ (referred as the Gel'fand-Dikii potential in \cite{aoyama:96})
  defined as the flux density of the conservation laws,
\begin{equation}
\label{G}
\frac{\partial g_j}{\partial t_i}=\frac{\partial}{\partial x}G_{ij}.
\end{equation}
Here
  $g_j=G_{0j}$ and $G_{ij}$ are the coefficients of the expansions of
$p_m$ and $Q_i^m$ in (\ref{pmflow}). In the case of the ${\rm BH}_m$, they
are given by the residue formulae,
\[
G_{ij}=\underset{k=\infty}{{\rm Res}}\left(Q_i^m~k^{j-m}~\frac{dk}{\sqrt{k}}
\right).
\]
Because of the conservation law (\ref{G}), $G_{ij}$ can be integrated once to
get the form with some function $F_j$,
\[
G_{ij}=\frac{\partial}{\partial t_i}F_j.
\]
(Note that $F_j$ are related with the functions $a_i$'s in (\ref{sm})
as $F_j=a_{m-1-j}$ for $ 0\le j\le m-1$.)
However one can show that $G_{ij}$ is not a symmetric function with respect
to the indices $i,j$ except the case with $m=0$. Hence there is no further
integration to express the $G_{ij}$ as a two point function as in the case of
dispersionless KP theory (see for example \cite{aoyama:96,kodama:98}), such as,
\[
G_{ij}=\langle \phi^i\phi^j\rangle =\frac{\partial^2}{\partial t_i\partial
t_j}F,
\]
where $\phi_i$'s are the primary fields or graviational decendants and $F$
represents the free energy of the topological field theory.
Thus the ${\rm BH}_m$ hierarchy is somewhat different from most of the known
dispersionless hierarchies which have a direct connection to topological
field theory \cite{witten:90,aoyama:96}.
It is then quite interesting to construct a topological field theory
based on the ${\rm BH}_m$ hierarchies, and if there exists such a theory,
then the associated Frobenius
manifold may have a stratified structure characterized by a finite ring
related to ${\mathcal F}_m$ in (\ref{F}).

\medskip
\noindent
{\bf Aknowledgement:~~}
YK is supported in part by NSF Grant \#DMS0071523.
BGK is supported in part by the grant COFIN 2000 \lq\lq Sintesi\rq\rq,
and is grateful to Department of Mathematics, Ohio State University for
the kind hospitality and support. BGK would like to thank L. Martinez
Alonso, A. Mikhailov
and A. Veselov for useful discussions.

\medskip
\noindent
{\bf Note added in proof:~~} After the completion of this paper, M. 
Pavlov informed us that the ${\rm BH}_m$ hierarchy can be transformed 
into
the Whitham equation correspoding to the $m$-gap solution of
the KdV equation by a {\it reciprocal transformation}. We thank M. Pavlov
for this information.

\bibliographystyle{amsalpha}

\end{document}